# Composition-Driven Tunable Optical and Electrical Properties in Van der Waals Ferroelectric $NbOI_{2-x}Cl_x$ Alloys

Gaolei Zhao,[1] Juhe Liu,[1, 2] Jinkai Huo,[1] Tian Han,[2] Yunhao Tong,[1] Hu Wang,[1] Konstantin Kozadaev,[3] Andrei Zheltkovich,[5] Changsen Sun,[1] Alexei Tolstik,[4] Andrey Novitsky,[3] Lujun Pan,[6] and Dawei Li[1, a)]

[1] School of Optoelectronic Engineering and Instrumentation Science, Dalian University of Technology, Dalian 116024, China

[2] Dalian University of Technology and Belarusian State University Joint Institute, Dalian University of Technology, Dalian, 116024, China

[3] Physical Optics and Applied Informatics Department, Belarusian State University, Minsk, 220030, Belarus

[4] Laser Physics and Spectroscopy Department, Belarusian State University, Minsk, 220030, Belarus

[5] Geotechnical Engineering and Construction Mechanics Department, Belarusian National Technical University, Minsk, 220013, Belarus

[6] School of Physics, Dalian University of Technology, Dalian 116024, China

[a)] Author to whom correspondence should be addressed: dwli@dlut.edu.cn

## ABSTRACT

Layered niobium oxide dihalides $NbOX_2$ (X = I, Cl), as a new family of Van der Waals (vdW) ferroelectrics, have attracted extensive attention, while achieving non-volatile modulation of their optical and electrical properties remains challenging, thereby limiting their integration into next-generation nanoelectronics and optoelectronics. Here, we report the controlled fabrication of highly crystalline $NbOI_{2-x}Cl_x$ vdW alloys with composition-driven tunable optical and electrical properties via a chemical vapor transport method. Comprehensive experimental characterization combined with first-principles calculation shows that the crystal lattices, phonon modes, and band structures of $NbOI_{2-x}Cl_x$ can be well tailored, which are distributed between $NbOI_2$ and $NbOCl_2$. Both the amplitude and polarization of second harmonic generation optical signal in $NbOI_{2-x}Cl_x$ exhibit pronounced compositional dependence, offering optical evidence for tunable in-plane ferroelectric characteristic. Moreover, field-effect transistors based on $NbOI_{2-x}Cl_x$ display robust *n*-type semiconducting behavior, with threshold voltage and carrier mobility precisely modulated through adjustment of I/Cl molar ratio. Furthermore, 2D $NbOI_{2-x}Cl_x$ photodetectors across all compositions exhibit exceptional gate-tunable current on/off ratio and strong polarization-sensitive photo-response. This study thus provides a new vdW ferroelectric material platform with tunable optical and electrical properties, paving the path for its implementation in modern nanophotonics and nanoelectronics.

In recent years, with the increasing demand in the miniaturization and functional integration of electronic devices, two-dimensional (2D) van der Waals (vdW) ferroelectrics have attracted significant attention due to their intrinsic scaling advantages.[1] So far, numerous 2D ferroelectrics have been theoretically predicted and experimentally confirmed,[2] such as $CuInP_2S_6$ with out-of-plane polarization,[3, 4] monochalcogenide MX (M = Sn, Ge, X = S, Te) with in-plane polarization,[5-7] and α-$In_2Se_3$ with correlated in-plane and out-of-plane polarization.[8, 9] More recently, layered $NbOX_2$ (X = Cl, I, Br), as a new family of 2D ferroelectric materials, has emerged.[10, 11] It has been shown that $NbOX_2$ possesses high spontaneous polarization,[12] large lateral piezoelectric coefficient,[10] weak interlayer coupling,[13, 14] strong nonlinear optical responses,[15-17] and highly anisotropic electrical, optical, thermoelectric, and mechanical properties.[18-20] These remarkable properties make $NbOX_2$ promising for various device applications, including field-effect transistors (FETs),[21, 22] quantum light sources,[13, 23] nonvolatile memories,[24, 25] sensors,[26, 27] and photodetectors.[28, 29]

To date, investigations into 2D $NbOX_2$ have primarily focused on device functionalities derived from its intrinsic ferroelectric, electrical, and optical properties, while achieving controllable modulation of these properties remains challenging, thereby limiting its broader potential in next-generation nanoelectronics and optoelectronics. Recently, several strategies have been proposed to modulate in-plane physical responses in 2D $NbOX_2$, including strain field,[30-32] high pressure field,[33, 34] temperature field,[35, 36] geometric parameter and vdW stacking engineering.[37, 38] However, most methods involve complex processes and rely on external energy field. A promising approach to non-volatile and sensitive engineering their physical properties is alloying.[39, 40] To the best of our knowledge, 2D ferroelectric $NbOX_2$-based alloys with tunable optical and electrical properties have not been explored to date.[41]

In this work, we report the controlled growth of vdW ferroelectric $NbOI_{2-x}Cl_x$ alloys across entire compositional range by a chemical vapor transport (CVT) strategy. A combined experimental characterization and first-principles calculation demonstrate that as-grown $NbOI_{2-x}Cl_x$ alloys are high-quality single crystals, whose lattice structures, phonon modes, and energy bandgaps distribute between $NbOI_2$ and $NbOCl_2$. In addition, the second-harmonic generation (SHG) nonlinear light intensity monotonically decreases with increasing Cl content, while the SHG light polarization pattern evolves from a four-lobed to a two-lobed configuration, confirming composition-driven tunability of in-plane ferroelectric character in $NbOI_{2-x}Cl_x$. By fabricating $NbOI_{2-x}Cl_x$ alloy FET devices, a continuous modulation of carrier mobility and threshold voltage is achieved through I/Cl molar ratio control. Furthermore, $NbOI_{2-x}Cl_x$ alloys working as photodetectors exhibit highly gate-tunable current on/off ratio and strong polarization-sensitive photo-response. Our study thus offers a new class of vdW ferroelectric material platform with highly tunable physical properties, paving the way for their applications in next-generation nanoelectronics and nanophotonics.

$NbOI_{2-x}Cl_x$ features a layered, low-symmetry monoclinic crystal structure within the space group *C*2 (Fig. 1a). Within this structure, Nb and O atoms form a planar [NbO] network, whereas I and Cl atoms are distributed between neighboring layers, coordinating with Nb to yield $[NbO_2I_{2x}Cl_{2(2-x)}]$ octahedra (Fig. 1b and 1c).[16, 18] Notably, Nb atoms deviate from the center of octahedra along both the *c*-axis and the *b*-axis. The *b*-axis displacement breaks inversion symmetry and generates a net spontaneous polarization (Fig. 1b), thereby establishing the intrinsic ferroelectric nature of $NbOI_{2-x}Cl_x$. High-quality $NbOI_{2-x}Cl_x$ alloys were grown via a CVT method (Fig. 1d). In brief, the mixed powder precursors with a molar ratio were sealed under vacuum in a quartz tube, and heated in a one-zone furnace at 700 °C and kept for 5 days (see supplementary materials). As shown in Fig. 1e, lustrous black bulk crystals with well-defined faceted morphologies were obtained across all compositions ($0 \leq x \leq 2$). To verify compositional fidelity, we carried out chemical analysis of $NbOI_{2-x}Cl_x$ samples in Fig. 1e by energy dispersive X-ray spectroscopy (EDX, Fig. 1f). The samples with $0 < x < 2$ exhibit clear signals from Nb, O, I, and Cl, confirming $NbOI_{2-x}Cl_x$ alloy formation. In contrast, the $x = 0$ and $x = 2$ end-members show only Nb, O, I and Nb, O, Cl, respectively, consistent with stoichiometric $NbOI_2$ and $NbOCl_2$. Fig. 1g compares the expected composition with the composition measured by EDX. There is a close correspondence between expected composition and actual composition, confirming that the stoichiometry and composition of $NbOI_{2-x}Cl_x$ can be well tunned through precursor mixture design.

To investigate the effect of composition on the structural property of $NbOI_{2-x}Cl_x$ alloys, X-ray diffraction (XRD) measurements were performed (Fig. 2a). All samples have four intensive and narrow diffraction peaks of (200), (400), (600), and (800), indicating high crystallinity and preferential (100) oriented growth. The peak position, such as (600) plane (Fig. 2b, upper panel), shift progressively with increasing Cl content, where a gradual increase is observed for $0 \leq x \leq 0.7$, followed by a more pronounced shift for $0.7 < x \leq 2$. Using Bragg equation ($d = \frac{n\lambda}{2\sin\theta}$), [42] the interlayer spacing can be extracted, where $\theta$ is the diffraction angle, $\lambda$=1.54 Å is the Cu-Ka X-ray wavelength, and $n$ is the diffraction order. Fig. 2b (lower panel) displays the dependence of crystal plane spacing on Cl composition. The interlayer spacing values span the entire range between those of pristine $NbOI_2$ and $NbOCl_2$, confirming composition-tunable lattice structure. For $NbOI_2$ and $NbOCl_2$, the extracted interlayer spacing is 0.73 nm and 0.62 nm, respectively, consistent with the previously reported results.[13, 18]

To study the phonon modes in $NbOI_{2-x}Cl_x$ alloys, we performed Raman measurements. Fig. 2c shows the Raman spectra for six $NbOI_{2-x}Cl_x$ alloys with different Cl compositions ($x$). The end point compounds of $NbOI_2$ ($x = 0$) and $NbOCl_2$ ($x = 2$) each display five distinct Raman-active peaks, in agreement with the previously reported data.[24, 28] Specifically, $NbOI_2$ exhibits peaks at ~102, ~206, ~272, ~511, and ~610 $cm^{-1}$ (labeled as $P_1$-$P_5$), whereas $NbOCl_2$ shows peaks at ~157, ~175, ~295, ~338, and ~668 $cm^{-1}$ (labeled as $P_1'$-$P_5'$). Notably, $NbOI_{2-x}Cl_x$ alloys ($0 < x < 2$) yield seven Raman peaks, including $P_1$, $P_1'$, $P_2$, $P_2'$, $P_3$ ($P_3'$), $P_4$ ($P_4'$), and $P_5$ ($P_5'$). It is suggested that phonon modes of $NbOI_{2-x}Cl_x$ alloy arise from hybridized contributions of $NbOI_2$ and $NbOCl_2$. Fig. 2d shows the Cl composition dependence of Raman peak position. All modes, except for $P_4$, undergo a blue shift as the Cl composition increases. This behavior can be explained that, in $NbOI_{2-x}Cl_x$ alloy, $P_4$ belongs to $B_g$-like symmetry mode, while the remaining modes possess $A_g$-like symmetry.[18] Among all peaks, $P_5$ ($P_5'$) mode exhibits the largest frequency shift per unit change in $x$, making it possible for quantifying the Cl/I stoichiometric ratio in $NbOI_{2-x}Cl_x$.

To determine the bandgaps of $NbOI_{2-x}Cl_x$ alloys, density functional theory calculations were performed. Two possible atomic configurations are considered plausible for $NbOI_{2-x}Cl_x$: (1) a configuration featuring halogen substitution on a single side of the layered structure (Supporting Information), and (2) a configuration with halogen atoms substituted symmetrically across both sides (Fig. 1a),[43] both of which yield nearly identical bandgap values. Fig. 2e-2g presents the calculated electronic band structures for three representative $NbOI_{2-x}Cl_x$. The conduction band minimum is predominately derived from Nb-4d states, whereas the valence band maximum arises from hybrid contributions of I-5p, Cl-5p, and Nb-4d orbitals.[18] Our calculation results demonstrate a monotonic increase in the bandgap, from 1.76 eV for $NbOI_2$ to 2.05 eV for $NbOCl_2$, indicating strong compositional tunability. In addition, the valence band maximum exhibits progressive flattening with increasing Cl composition, reflecting enhanced effective mass and reduced carrier mobility along this band edge.

SHG is one of the most important nonlinear optical phenomena in 2D ferroelectrics.[13, 15] We focused a picosecond laser ($\lambda$ = 1064 nm) onto the sample surface at normal incidence and collected the SHG signal of $NbOI_{2-x}Cl_x$ alloys in reflection mode (Fig. 3a). Fig. 3b presents an optical image of a typical $NbOI_{2-x}Cl_x$ ($x$ = 0.9) alloy flake on $SiO_2$/Si substrate. The corresponding SHG intensity mapping (Fig. 3c) reveals pronounced thickness-dependent contrast, indicating strong modulation of the nonlinear optical response via layer thickness. Quantitative analysis (Fig. 3d) shows that the SHG signal increases sharply with thickness up to ~30 nm, followed by a gradual decrement at larger thicknesses. This thickness scaling behavior is similar to that reported for other 2D ferroelectrics, including $NbOCl_2$ and $CuInP_2S_6$.[13, 44] Fig. 3e displays the SHG intensity as a function of incident laser power measured from a $NbOI_{1.1}Cl_{0.9}$ flake (inset). A linear fit to the logarithmic plot yields an exponent of ~1.9, further confirming that the collected signal stems from the SHG emission.

Next, we examine the influence of composition variation on the SHG response in $NbOI_{2-x}Cl_x$. Fig. 3f shows the SHG spectra of $NbOI_{2-x}Cl_x$ with comparable layer thickness but varying Cl content ($x$). The normalized SHG intensity, defined as SHG intensity per unit layer thickness, as a function of the Cl component ($x$) is shown Fig. 3g. A pronounced monotonic decrease in SHG intensity is observed with increasing $x$, achieving over one order-of-magnitude suppression between $x = 0$ and $x = 2$. Given that the second-order nonlinear response in $NbOX_2$ compound is intrinsically linked to their in-plane ferroelectric polarization, compositional tuning strategy provides an effective means to modulate ferroelectric behavior in layered niobium oxyhalides.

To investigate the polar symmetry of $NbOI_{2-x}Cl_x$, we performed polarization-resolved SHG measurements. Polarized SHG signals were collected under parallel and perpendicular polarization configurations by rotating a half-wave plate from $\Phi = 0^o$ to 180°. Fig. 3h-3k and S4 compare the polarized SHG responses for few-layer $NbOI_{2-x}Cl_x$ with varying Cl compositions. In the perpendicular configuration, all compositions exhibit a highly anisotropic two-lobed SHG pattern. In contrast, the parallel configuration yields a four-lobed pattern for $0 \leq x \leq 0.2$, featuring primary maximum at $\theta \approx$ 50°, 130°, 230°, and 310°, and second maximum at $\theta \approx$ 90° and 270°. Upon increasing the Cl composition beyond $x$ = 0.2 (up to $x$ = 2), the pattern progressively evolves into a two-lobed configuration, which intensity maximum shifts to $\theta \approx$ 90° and 270°. These observations indicate that the SHG light polarization in $NbOI_{2-x}Cl_x$ can be effectively tuned via precise control of the I/Cl molar ratio.

The polarized SHG intensity can be expressed as: [16]

$$I_{\mathrm{SHG}} \propto \left| \hat{e}_{2\omega} d^{(2)}_{\mathrm{NbOI}_{2-x}\mathrm{Cl}_x} \hat{e}^2_{\omega} \right|^2, \quad (1)$$

where $\hat{e}_{\omega} = [\cos(\theta), -\sin(\theta), 0]$, $\hat{e}_{2\omega} = \begin{cases} [\cos(\theta), -\sin(\theta), 0] \ (\hat{e}_{2\omega} \parallel \hat{e}_{\omega}) \\ [\sin(\theta), \cos(\theta), 0] \ (\hat{e}_{2\omega} \perp \hat{e}_{\omega}) \end{cases}$, and $d^{(2)}_{\mathrm{NbOI}_{2-x}\mathrm{Cl}_x}$ is the second-order $d$-tensor. Here, $\theta$ represents the angle between the incident light polarization and the $x$-axis. The $d$-tensor can be written as:[15]

$$d_{\mathrm{NbOI}_{2-x}\mathrm{Cl}_x} = \begin{pmatrix} 0 & 0 & 0 & d_{14} & 0 & d_{16} \\ d_{21} & d_{22} & d_{23} & 0 & d_{25} & 0 \\ 0 & 0 & 0 & d_{34} & 0 & d_{36} \end{pmatrix}, \quad (2)$$

where $d_{16} = d_{21}$, $d_{14} = d_{25} = d_{36}$, $d_{23} = d_{34}$. Substituting this tensor into equation (1), the SHG intensity expressions under parallel and perpendicular configurations are derived as follows :

$$I_{\parallel,\mathrm{SHG}} \propto d_{22}^2 \left| (3\frac{d_{21}}{d_{22}} \sin(\theta) \cos^2(\theta) + \sin^3(\theta)) \right|^2, \quad (3)$$

$$I_{\perp,\mathrm{SHG}} \propto d_{22}^2 \left| (1 - 2\frac{d_{21}}{d_{22}}) \cos(\theta) \sin^2(\theta) + \frac{d_{21}}{d_{22}} \cos^3(\theta) \right|^2. \quad (4)$$

The total SHG intensity is obtained by summing up $I_{\parallel,\mathrm{SHG}}$ and $I_{\perp,\mathrm{SHG}}$:

$$I_{\mathrm{total}} \propto d_{22}^2 ((\frac{d_{21}}{d_{22}} \sin(2\theta))^2 + (\frac{d_{21}}{d_{22}} \cos^2(\theta) + \sin^2(\theta))^2). \quad (5)$$

The above electromagnetic theory indicates that polarization-dependent SHG response in $NbOI_{2-x}Cl_x$ is predominantly governed by the relative magnitude of two tensor elements, $d_{21}$ and $d_{22}$. Fitting polarization-resolved SHG data using equations (3) and (4) yields excellent agreement. Crucially, the evolution of SHG polarization (Fig. 3h-3k) and intensity (Fig. 3f) across the $NbOI_{2-x}Cl_x$ series is quantitatively captured by the dimensionless ratio $\frac{d_{21}}{d_{22}}$ (Fig. 3g, right panel), which decreases monotonically from 0.8 ($x$ = 0) to 0.2 ($x$ = 2).

Next, we fabricated FET devices based on $NbOI_{2-x}Cl_x$ alloys and systemically explored the possibility of threshold voltage engineering and carrier mobility modulation, which is critical for their integration into 2D CMOS logic circuits.[45] In our device (Fig. 4a), a multilayer $NbOI_{2-x}Cl_x$ flake serves as the semiconducting channel, a 300-nm-thick $SiO_2$ layer works as the bottom gate dielectric, while few-layer graphene electrodes provide source and drain contacts. During device fabrication, the non-polar *c*-axis of $NbOI_{2-x}Cl_x$ is aligned parallel to two electrodes. Fig. 4b presents an optical image of a multilayer (~27 nm) $NbOI_{1.1}Cl_{0.9}$ FET device. Atomic force microscopy analysis confirms an atomically smooth surface morphology (Fig. 4c). Fig. 4d displays the output ($I_d$-$V_d$) characteristics of the device taken under varying gate voltages ($V_g$). As $V_g$ increases from -40 to +40 V, the channel conductance is progressively enhanced, confirming an *n*-type semiconducting characteristic of $NbOI_{2-x}Cl_x$ alloy. The transfer ($I_d$-$V_g$) characteristics measured on the same device are shown in Fig. 4e. At $V_d$ = 0.1V, the device achieves a current on/off ratio exceeding $2.4 \times 10^5$, demonstrating excellent switching performance and great potential for low-power logic applications.

In Fig. 4f, we compare the transfer characteristics of FET devices fabricated using 2D $NbOI_{2-x}Cl_x$ with varying Cl compositions (Supporting Information). As shown in Fig. 4g (left panel), as the composition in $NbOI_{2-x}Cl_x$ evolves from $NbOI_2$ to $NbOCl_2$, the extracted $V_{th}$ varies over a broad range from +20.1 V to -34.5 V. This composition-driven wide-range modulation of $V_{th}$ enables selective design of enhancement- and depletion-mode FET devices, simply by tailoring the Cl stoichiometry in $NbOI_{2-x}Cl_x$. We also extracted the field effect mobility ($\mu_{EF}$) from the linear region of $I_d$-$V_g$ curves as $\mu_{EF} = \frac{1}{C_{SiO_2}} \cdot \frac{L}{W} \cdot \frac{1}{V_d} \cdot \frac{dI_d}{dV_g}$,[46, 47] where $C_{SiO_2} = \frac{\varepsilon_o \varepsilon_r}{d}$ is the capacitance of $SiO_2$ and $L$ ($W$) is the channel length (width). As illustrated in Fig. 4g (right panel), $\mu_{EF}$ reaches the maximum (~1.63 $cm^2V^{-1}s^{-1}$) at $x = 0$ and decreases monotonically to 0.02 $cm^2V^{-1}s^{-1}$ at $x = 2$. The exponential decrease in $\mu_{EF}$ with increasing Cl content is attributed to Cl doping-induced carrier scattering and carrier relaxation effect.[48]

Finally, we investigate the optoelectrical performance of $NbOI_{2-x}Cl_x$ alloy devices (Fig. 5a-5c). Fig. S6 shows the optical images of three multilayer $NbOI_{2-x}Cl_x$ photodetectors with Cl composition of $x$ = 0, 0.9, and 2, respectively. $I_d$-$V_d$ measurements ($V_g$ = 0 V) under dark and illumination reveal a composition-dependent photo-response, where the current on-off ratio ($I_{ph}/I_{dark}$) monotonically decreases with increasing Cl content. This behavior is primarily attributed to Cl doping-induced dark current enhancement. The gate-voltage-dependent $I_d$-$V_g$ characteristics under dark and illumination (Fig. 5d-5f, left panel) demonstrate robust gate tunability of the photo-response. To quantitatively evaluate gate modulation, we plot $I_{ph}/I_{dark}$ ratio as a function of $V_g$ (Fig. 5d-5f, blue dotted lines). The pristine $NbOI_2$ device ($x$ = 0) maintains a high switching ratio (>$10^6$) over an extended gate-voltage window (-40 V≤ $V_g$ ≤ 20 V), reaching a maximum value of $3\times10^6$ at $V_g$ = 16.5 V (Fig. 5d). In contrast, the effective gating window for high on/off photo-response becomes narrow with increasing $x$, consistent with Cl-composition-dependent shift in the threshold voltage.

Time-resolved photo-response measurements (Fig. 5g) indicate that all $NbOI_{2-x}Cl_x$ devices exhibit second-scale rise and fall times. As $x$ increases from 0 to 2, both the rising time and decaying time increase by approximately two orders of magnitude (Fig. 5i, left panel). This pronounced slowdown is ascribed to reduced carrier mobility and enhanced trapping states introduced by Cl substitution.[49, 50] Subsequently, we carry out polarization-resolved photo-response measurements (Fig. 5h). The photocurrent of $NbOI_{2-x}Cl_x$ device is maximized when the incident light polarization aligns parallel to $c$-axis, consistent with its pristine single crystal.[28] We define the anisotropic photo-response factor: $F_{ph} = I_{ph}^{b}/I_{ph}^{c}$, and find that $F_{ph}$ decreases linearly from 2.6 ($NbOI_2$) to 1.7 ($NbOCl_2$). Overall, our study establishes that both electrical and optoelectronic properties $NbOI_{2-x}Cl_x$ alloys can be precisely engineered by simply adjusting I/Cl molar ratio.

In summary, we have successfully achieved single-crystalline, 2D ferroelectric $NbOI_{2-x}Cl_x$ alloy semiconductors. By using a precursor mixture of $Nb_2O_5$, Nb, $I_2$, $NbCl_5$, our CVT method yields $NbOI_{2-x}Cl_x$ alloys whose compositions match closely those expected from the nominal I/Cl molar ratios. Comprehensive characterization and theoretical calculations confirm composition-dependent modulation of lattice structures, phonon modes, and electronic bandgaps across the $NbOI_{2-x}Cl_x$ alloy series. Nonlinear optical measurements reveal tunable in-plane ferroelectricity and optical anisotropy as a function of Cl content. Electrical transport studies demonstrate that threshold voltage can be precisely engineered through compositional control, enabling selective design of enhancement- and depletion-mode FET devices. Moreover, $NbOI_{2-x}Cl_x$ alloy photodetectors exhibit exceptional gate-voltage and polarization sensitivity. Our work thus offers a new vdW ferroelectric material platform featuring highly tunable structural, optical and electrical properties.

This work was supported by the National Natural Science Foundation of China (Grant No. 12274051), the Natural Science Foundation of Liaoning Province (Grant No. 2024-MSBA-06), the Liaoning Province Xingliao Talents Plan Project (Grant No. XLYC2403069), the Fundamental Research Funds for the Central Universities (Grant Nos. DUT24RC(3)060, DUT22ZK109), the High-end Foreign Experts Recruitment Plan of China (Grant No. gz120250047), the National Key Research and Development Program of China (Grant No. 2024YFE0213500), and the Project of the China Scholarship Council to Promote International Cooperation and Training with Russia, Ukraine and Belarus (Grant No. EWXM2509110228).

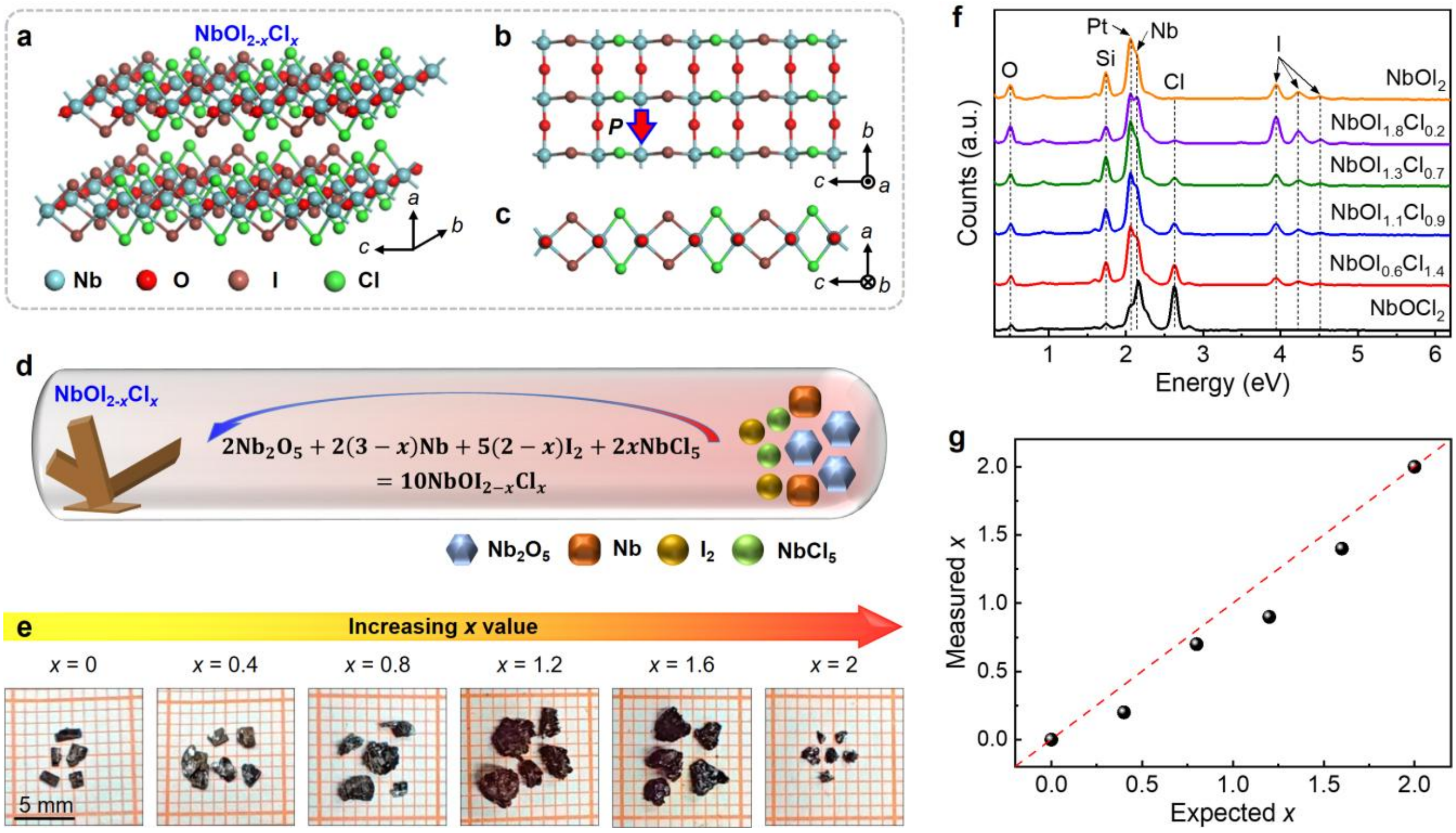

**FIG. 1.** Schematics of the crystal structure of 2D $NbOI_{2-x}Cl_x$: (a) 3D view, (b) top view, (c) and side view. (d) Schematic of the experimental setup for $NbOI_{2-x}Cl_x$ growth. (e) Optical photography of as-grown single-crystalline $NbOI_{2-x}Cl_x$ bulk samples with different compositions. (f) EDX spectra acquired from $NbOI_{2-x}Cl_x$ samples shown in (e). (g) Plot of expected compositions and the corresponding compositions measured by EDX.

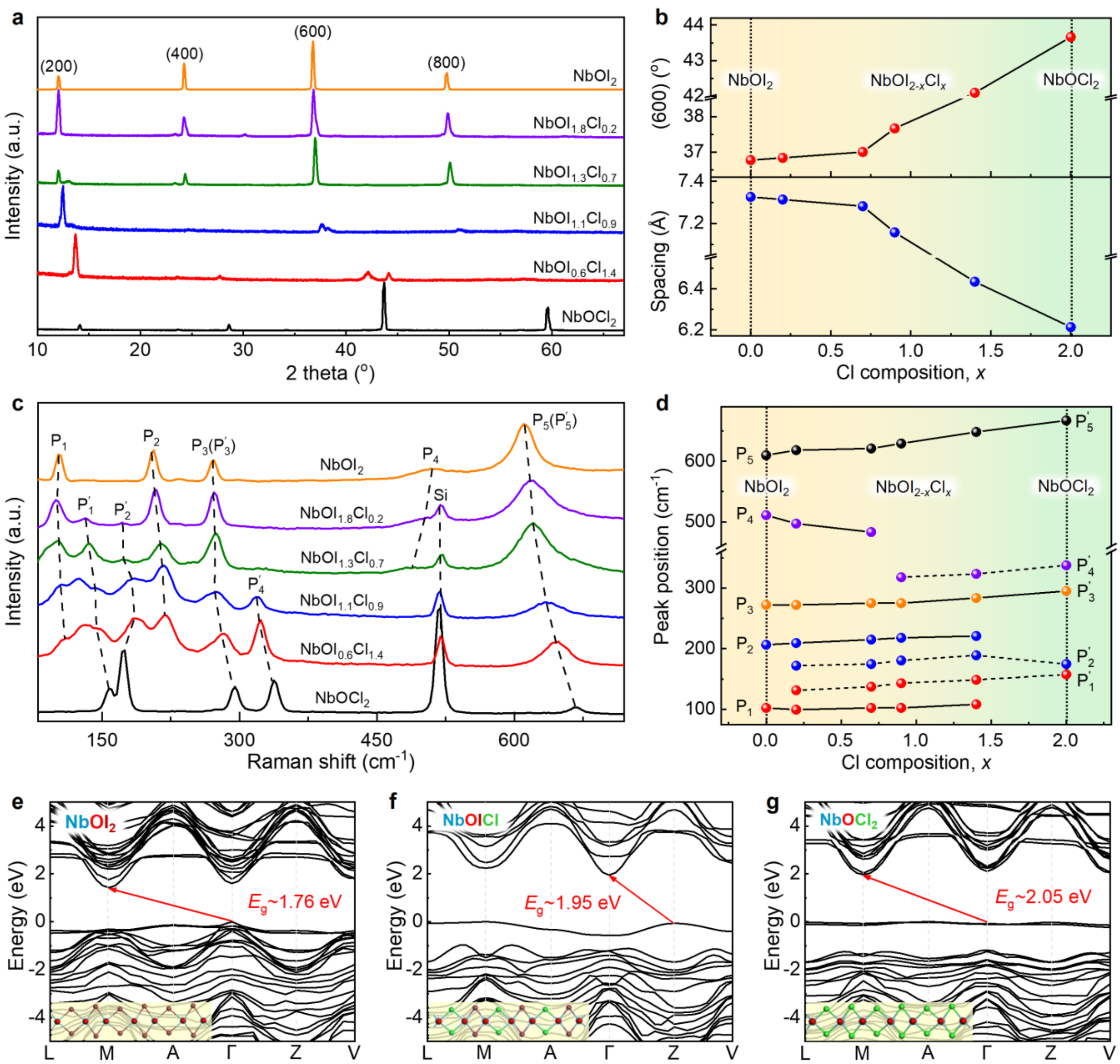

**FIG. 2.** (a) XRD $2\theta/\omega$ scan of six $NbOI_{2-x}Cl_x$ samples spanning the full compositional range ($0 \leq x \leq 2$). (b) (600) diffraction peak position (upper panel) and corresponding crystal plane spacing (lower panel) as a function of Cl composition, $x$. (c) Raman spectra of six $NbOI_{2-x}Cl_x$ alloys with different compositions. (d) The dependence of selected Raman-active mode frequency on $x$, extracted from the data in (c). (e-g) Calculated electronic band structures for (e) $NbOI_2$, (f) NbOICl, and (g) $NbOCl_2$.

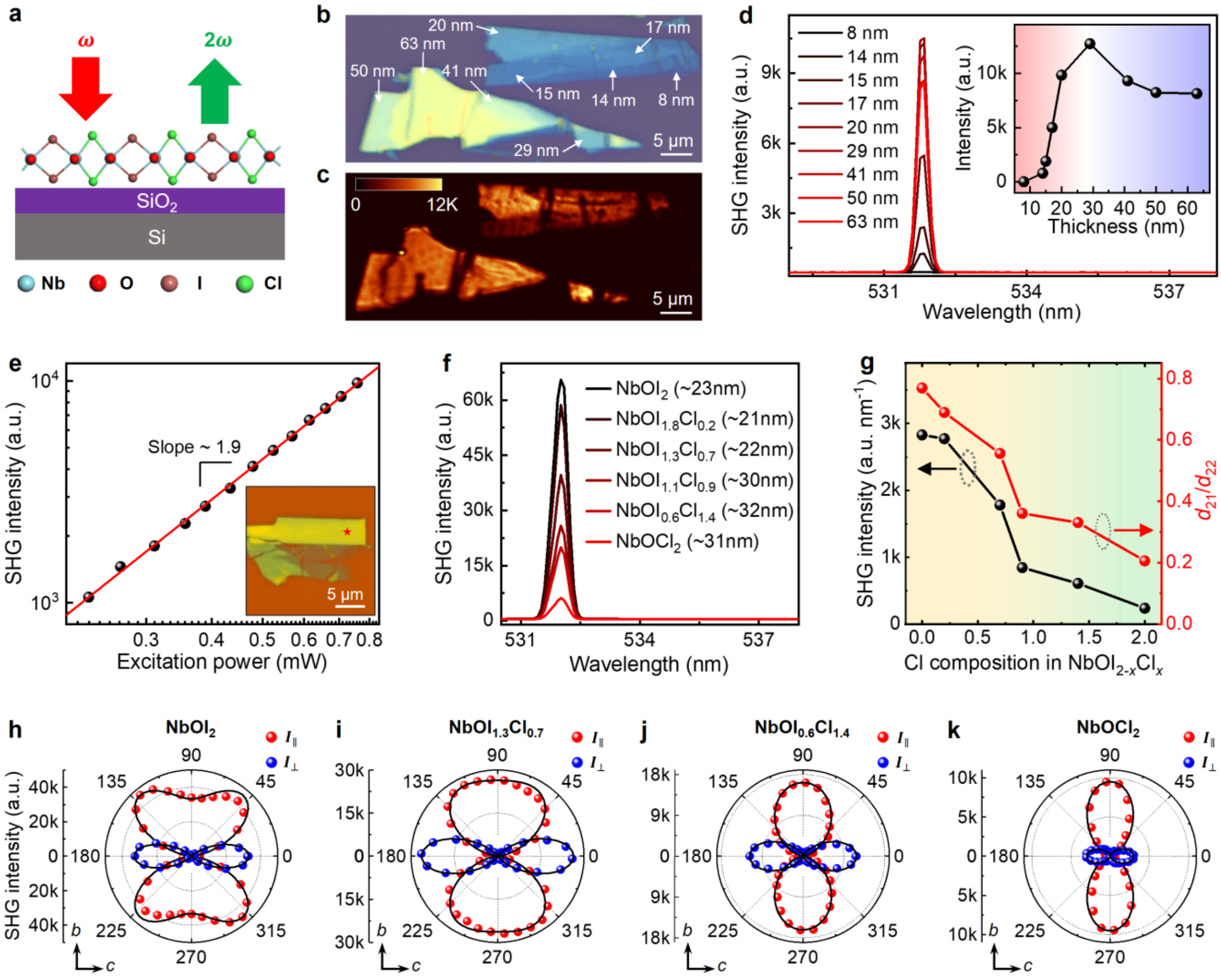

**FIG. 3.** (a) Schematic of the SHG experimental setup. (b) Optical image and (c) SHG intensity mapping of an exfoliated $NbOI_{1.1}Cl_{0.9}$ flake with different layer thicknesses. (d) SHG spectra acquired from $NbOI_{1.1}Cl_{0.9}$ flakes in (b). Inset: SHG intensity as a function of layer thickness. (e) Excitation-power-dependent SHG intensity measured from a $NbOI_{1.1}Cl_{0.9}$ flake (inset). (f) SHG spectra of $NbOI_{2-x}Cl_x$ alloys of similar layer thickness with varied Cl composition. (g) SHG intensity and $d_{21}/d_{22}$ ratio as a function of Cl component $x$ in $NbOI_{2-x}Cl_x$. (h-k) Polar plots of parallel (red dots) and perpendicular (blue dots) polarized SHG intensity as a function of the sample angle $\theta$ for (h) $NbOI_2$, (i) $NbOI_{1.3}Cl_{0.7}$, (j) $NbOI_{0.6}Cl_{1.4}$, and (k) $NbOCl_2$.

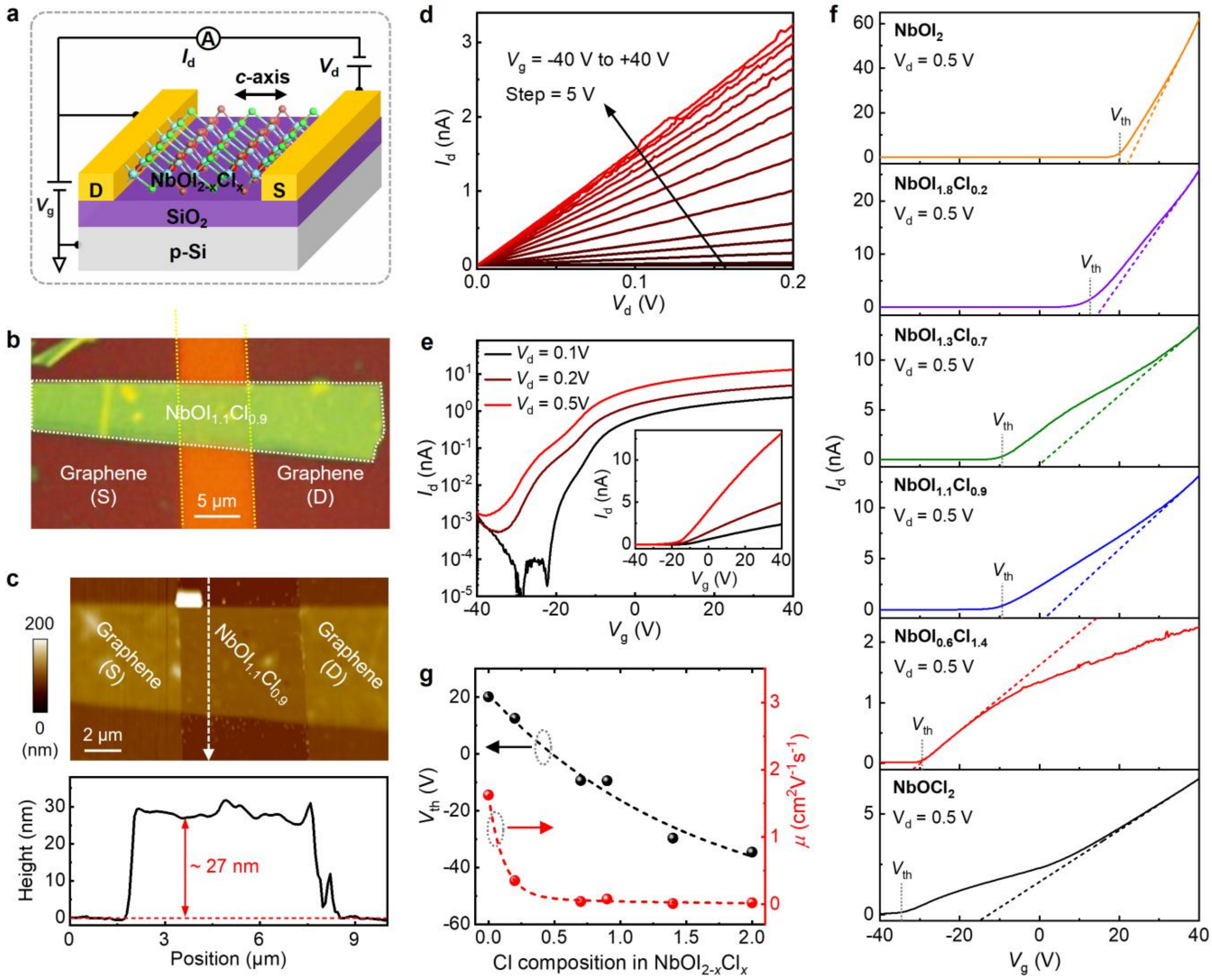

**FIG. 4.** (a) Schematic of the $NbOI_{2-x}Cl_x$ FET device. (b) Optical image and (c) AFM topography of a multilayer (~27 nm) $NbOI_{1.1}Cl_{0.9}$ device. (d) Output ($I_d$-$V_d$) characteristics of the same device in (b) taken at various $V_g$. (e) Transfer ($I_d$-$V_g$) characteristics of the same device in (b) taken at different $V_d$. (f) $I_d$-$V_g$ characteristics for multilayer $NbOI_{2-x}Cl_x$ alloy FETs with varied Cl compositions. (g) Extracted $V_{th}$ (left) and $\mu$ (right) as a function of Cl composition in $NbOI_{2-x}Cl_x$.

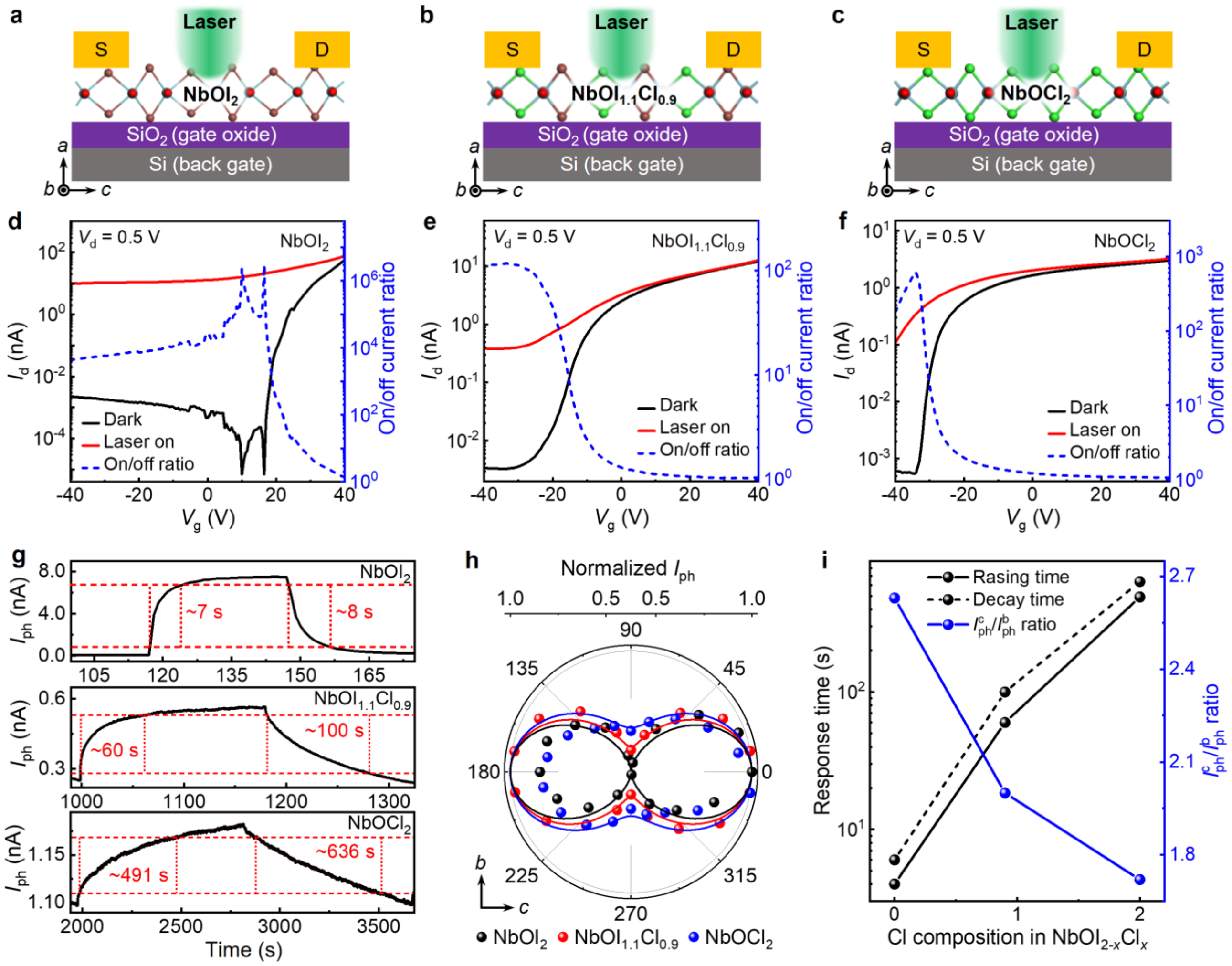

**FIG. 5.** Schematics of the $NbOI_{2-x}Cl_x$ alloy photodetectors: (a) $NbOI_2$, (b) $NbOI_{1.1}Cl_{0.9}$, and (c) $NbOCl_2$. $I_d$-$V_g$ ($V_d = 0.5V$) curves measured under dark and laser illumination (left) and photocurrent on/off ratio (right) for multilayer (d) $NbOI_2$, (e) $NbOI_{1.1}Cl_{0.9}$, and (f) $NbOCl_2$ devices. (g) Photo-response time measurement for devices in (d-f). (h) Polar plots of normalized photocurrent $I_{ph}$ ($V_g$ = 0 V) as a function of incident light polarization angle $\Phi$ for $NbOI_{2-x}Cl_x$ devices. (i) Cl composition-dependent photo-response time (left) and $I_{ph}^{c}/I_{ph}^{b}$ ratio (right).